Spontaneously generated waves and obstacles to integrability in perturbed evolution equations

Alex Veksler[1] and Yair Zarmi[1,2]

Ben-Gurion University of the Negev, Israel

[1]Department of Physics, Beer-Sheva, 84105

[2]Department of Energy & Environmental Physics

Jacob Blaustein Institutes for Desert Research, Sede-Boqer Campus, 84990

Abstract

The perturbed Burgers and KdV equations are considered. Often, the perturbation excites waves that are different from the solution one is seeking. In the case of the Burgers equation, the spontaneously generated wave is also a solution of the equation. In contrast, in the case of the KdV equation, this wave is constructed from new (non-KdV) solitons that undergo an elastic collision around the origin. Their amplitudes have opposite signs, which they exchange upon collision. The perturbation then contains terms, which represent coupling between the solution and these spontaneously generated waves. Whereas the unperturbed equations describe gradient systems, these coupling terms may be non-gradient. In that case, they turn out to be obstacles to asymptotic integrability, encountered in the analysis of the solutions of the perturbed evolution equations.

PACS 02.30.IK, 02.30.Mv, 05.45.-a

Key words: Perturbed integrable systems; Obstacles to integrability; Wave interactions; KdV equation; Burgers equation.

Integrable nonlinear evolution equations are usually derived as approximations to more complex physical systems. Of this type are, for example, the Burgers equation (propagation of weak shock fronts in a fluid [1, 2]), the KdV equation (propagation of solitons in shallow water [3, 4]), and the NLS equation (propagation of electromagnetic pulses in optical fibers [5, 6]). When physically relevant higher-order terms are added to the equation, the ability to generate a perturbative expansion for the solution depends on the type of solution sought [7-12].

If one is seeking a solution, for which the zero-order approximation is a single-wave (front or soliton), then the standard Normal Form expansion can be implemented. The Normal Form (NF, the equation governing the evolution of the zero-order approximation to the solution) is integrable through every order. Its solution has the single-wave structure of the unperturbed equation, except for a necessary update of the wave velocity. The Near-Identity Transformation (NIT, the expansion of the full solution in a power series in a small parameter) can be computed order-by-order in closed form as a functional of the zero-order approximation, and only contains bounded corrections.

The situation is different in the case of a multiple-wave solution (i.e., multiple solitons in cases of the KdV and NLS equations [8-10], or multiple fronts in case of the Burgers equation [7, 11, 12]). The perturbation contains terms that cannot be accounted for in the standard way. Trying to account for these terms by the NF, they spoil its integrability, and the resulting zero-order approximation loses the simple multiple-wave structure of the solution of the unperturbed equation. For this reason, the unaccounted-for terms are called "obstacles to asymptotic integrability". Alternatively, the effect of the obstacles on the higher-order corrections in the NIT cannot be expressed in closed form, e.g., as differential polynomials in the zero-order approximation. Moreover, the obstacles may lead to the appearance of unbounded terms in the perturbative expansion of the full solution. Ways to overcome the obstacle problem in the cases of the perturbed Burgers and KdV equations have been proposed in [13-15].

In [10], it has been suggested that obstacles to asymptotic integrability emerge owing to inelastic interactions. This note focuses on the physical significance of the obstacles in the Burgers and KdV equations. We show how the obstacles are the expression of inelastic interactions, stressing the following points:

*(i)*    The perturbation excites new waves, different from the solution one is seeking;

*(ii)*    The perturbation contains terms that represent coupling between the solution sought and these spontaneously generated waves;

*(iii)*   Whereas the unperturbed equations describe gradient systems, these coupling terms may be non-gradient. In that case, they turn out to be obstacles to asymptotic integrability;

*(iv)*   Owing to the dissipative nature of the Burgers equation, an infinitely extended overlap between the solution one is seeking and the spontaneously generated wave exists. Because of this overlap, the obstacle has the capacity to generate unbounded terms in the solution;

*(v)*   Owing to the conservative nature of the KdV equation, the overlap between the multiple-soliton solution and the spontaneously generated wave is confined to a local neighborhood of the origin. As a result, the obstacle generates an asymptotically decaying exponential tail.

The solution of the unperturbed equations can be constructed from solutions of a member of a Lax pair [16]. In the case of the KdV equation, the *N*-soliton solution is constructed from *N* Jost functions that correspond to the discrete spectrum of the Schrödinger equation [17-20]. The multiple-soliton solution is a specific combination of these functions [21]. Far from the origin, each Jost function generates one soliton with the wave number and velocity of the function. The perturbation contains terms, which represent coupling between the solution one is seeking, and another wave. The latter is *not* a solution of the KdV equation, because it is constructed through inelastic interactions amongst pairs of the same Jost functions. In second order, this coupling term is the obstacle to integrability.

In the case of the Burgers equation, the multiple-front solution is constructed from exponential waves that are solutions of the diffusion equation [22-24]. Focusing on front solutions that vanish in some direction in the *x-t* plane, we show that, again, the obstacle to integrability represents coupling between such fronts and other wave fronts. The latter are also solutions of the Burgers equation, however, *not* of the same family, because they obey different boundary conditions. In addition, they are generated through inelastic interactions of the fundamental exponential waves.

Consider, first, the perturbed Burgers equation, where an obstacle to integrability emerges already in the first-order calculation [7, 11, 12]:

$$w_t = 2 w w_x + w_{xx} + \varepsilon \left( 3\alpha_1 w^2 w_x + 3\alpha_2 w w_{xx} + 3\alpha_3 w_x^2 + \alpha_4 w_{xxx} \right) \ . \tag{1}$$

In the following, we shall need $S_n$, the symmetries of the Burgers equation, as well as their spatial integrals, $G_n$ [22-24]. In a first-order analysis, the first three symmetries are needed:

$$S_1[w] = w_x, \qquad G_1[w] = w$$
$$S_2[w] = 2ww_x + w_{xx}, \qquad G_2[w] = w^2 + w_x \quad . \tag{2}$$
$$S_3[w] = 3w^2 w_x + 3ww_{xx} + 3w_x^2 + w_{xxx}$$

Note that, whereas the unperturbed part of the r.h.s. of Eq. (1) can be written as a gradient term, namely, as a derivative with respect to $x$ of a differential polynomial in $w$, the perturbation cannot. Its decomposition into gradient and non-gradient terms is not unique. Based on [14, 15] we use the following decomposition:

$$3\alpha_1 w^2 w_x + 3\alpha_2 ww_{xx} + 3\alpha_3 w_x^2 + \alpha_4 w_{xxx} =$$
$$\left(\alpha_1 - \tfrac{1}{2}\alpha_2 + \tfrac{1}{2}\alpha_3\right) S_3[w] + \left(\tfrac{1}{2}\alpha_1 - \alpha_2 - \tfrac{1}{2}\alpha_3 + \alpha_4\right)\partial_x^2 S_1[w] + \left\{-\tfrac{3}{2}\alpha_1 + \tfrac{3}{2}\alpha_2\right\}\partial_x S_2[w] \quad . \tag{3}$$
$$+ \left(\tfrac{3}{2}\alpha_2 - \tfrac{3}{2}\alpha_3\right) R_{21}[w]$$

$R_{21}[w]$ is expressed in terms of symmetries of the Burgers equation [14, 15]:

$$R_{21}[u] = S_2[u] G_1[u] - S_1[u] G_2[u] \quad . \tag{4}$$

Unlike the first three terms on the r.h.s. of Eq. (3), it is not a gradient term.

Through $O(\varepsilon)$, the Near Identity Transformation (NIT) is written as

$$w = u + \varepsilon u^{(1)} + O\left(\varepsilon^2\right) \quad . \tag{5}$$

The first-order correction is written as

$$u^{(1)} = au^2 + cu_x + w^{(1)}(t, x) \tag{6}$$

(When $u$ is a single-front solution, another differential monomial, $u_x \partial_x^{-1} u$, is allowed in Eq. (6). However, it is not allowed in the multiple-front case, as there it generates linearly unbounded terms in the first-order correction [15].) The last term in Eq. (6) represents a contribution that cannot be written as a differential polynomial in $u(t,x)$.

The Normal Form (NF) is written as

$$u_t = S_2[u] + \varepsilon U_1[u] + O\left(\varepsilon^2\right) \quad . \tag{7}$$

Substituting Eqs. (5), (6) and (7) in Eq. (1), one obtains the first-order *homological equation*:

$$U_1 + \partial_t w^{(1)} - 2\partial_x\left(u\,w^{(1)}\right) - \partial_x^2 w^{(1)} = \mu S_3[u] + \rho\,\partial_x^2 S_1[u] + \sigma\,\partial_x S_2[u] + \lambda R_{21}[u] \quad. \tag{8}$$

In Eq. (8),

$$\mu = a + \alpha_1 - \tfrac{1}{2}\alpha_2 + \tfrac{1}{2}\alpha_3 \quad, \tag{9}$$

$$\lambda = a + \tfrac{3}{2}\alpha_2 - \tfrac{3}{2}\alpha_3 \quad, \tag{10}$$

$$\rho = \tfrac{1}{2}\alpha_1 - \alpha_2 - \tfrac{1}{2}\alpha_3 + \alpha_4 \quad, \tag{11}$$

$$\sigma = a + \tfrac{3}{2}\alpha_1 - \tfrac{3}{2}\alpha_2 \quad. \tag{12}$$

The symmetry on the r.h.s. of Eq. (8) is accounted for by its customary assignment to the NF:

$$U_1 = \mu S_3[u] \quad. \tag{13}$$

The next two terms generate bounded contributions in $w^{(1)}$ [14, 15].

Of the four terms on the r.h.s. of Eq. (8), only the last one is not a pure gradient term. If accounted for by the NF (i.e., added to the first-order term in Eq. (3)), it spoils the integrability of the latter [7, 11, 12]. This is the origin of the name "obstacle to asymptotic integrability" [7-12]. If one attempts to account for it by the first-order term in Eq. (2), then it has the capacity to generate a secular term in $u^{(1)}$. To avoid this, one chooses the coefficient $a$ so that $\lambda = 0$ in Eq. (8) [14, 15].

To expose the physical nature of the obstacle (Eq. (4)), we focus on multiple-wave solutions of the NF that vanish at infinity in some direction in the $x$-$t$ plane, given by

$$u(t,x) = \frac{\displaystyle\sum_{i=1}^{M} k_i\, e^{k_i\left(x + v_i t + x_{i,0}\right)}}{1 + \displaystyle\sum_{i=1}^{M} e^{k_i\left(x + v_i t + x_{i,0}\right)}} \quad , \quad v_i = k_i + \varepsilon\mu k_i^{2} + O\!\left(\varepsilon^{2}\right) \quad. \tag{14}$$

(Solutions that do not vanish at infinity in any direction in the plane can be transformed into ones that do vanish asymptotically in some direction by an appropriate Galilean transformation.)

In the case of a single-wave ($M = 1$), the solution of the NF is:

$$u(t,x) = \frac{k\, e^{k(x + vt + x_0)}}{1 + e^{k(x + vt + x_0)}} \quad , \quad v = k + \varepsilon\alpha_4 k^{2} + O\!\left(\varepsilon^{2}\right) \quad. \tag{15}$$

For this solution, the obstacle of Eq. (4) vanishes explicitly, because then the symmetries obey

$$S_n = k^{n-1} S_1 \quad, \quad G_n = k^{n-1} G_1 \quad. \tag{16}$$

For $M > 1$, the solution is composed of $M + 1$ semi-infinite fronts, which, away from the origin, asymptote into single-front solutions of the Burgers equation [26, 27, 15]. Two of the fronts are generated, each by one exponential wave. For instance, if one has $0 < k_1 < k_2 < \cdots < k_M$, then two fronts asymptote to single fronts that vanish at $x = -\infty$. They are generated each from one of the exponential waves with $i = 1, M$, and are given by Eq. (15). These two fronts are unaffected by the existence of other exponential waves. The remaining $(M-1)$ fronts are affected by interactions of the exponential waves. Each has two non-vanishing boundary values and is generated by inelastic interaction between pairs of exponential waves with *adjacent* wave numbers. Their asymptotic expression is given by [26, 27, 15]:

$$\frac{k_j + k_{j+1} \, e^{(k_{j+1} - k_j)(x + v_{j,j+1} t + \xi_j)}}{1 + e^{(k_{j+1} - k_j)(x + v_{j,j+1} t + \xi_j)}} \,, \qquad v_{j,j+1} = k_j + k_{j+1} + O(\varepsilon), \qquad 1 \le j \le M - 1 \quad. \tag{17}$$

The obstacle, given by Eq. (4), does not vanish for the $M > 1$ solution, because Eq. (16) is not obeyed then. Thus, this "canonical" obstacle represents the net effect of the multiple-front solution relative to the single-front one. (Because of the freedom inherent in the expansion, various forms for the obstacles to integrability can be found in the literature. However, they do not vanish explicitly when computed for the single-front case, unless constructed from the canonical obstacle.) We rewrite $R_{21}$ as:

$$R_{21} = u^2 \, \partial_x u_s \,, \qquad \left( u_s(t,x) \equiv \frac{G_2}{G_1} \right) \tag{18}$$

$u_s$ is constructed from the same $M$ exponential waves from which the zero-order term, $u$, is constructed (see Eq. (14)). Direct substitution shows that if $u$ is a solution of the NF, Eq. (7), then so is $u_s$. However, the two waves are different. $u$ has one vanishing boundary value, and the $(M + 1)$ semi-infinite fronts described above, whereas the new wave has only non-zero boundary values and $M$ semi-infinite fronts, all generated inelastically. Consider the two-wave case:

$$u(t,x) = \frac{k_1 \, e^{k_1 (x + v_1 t)} + k_2 \, e^{k_2 (x + v_2 t)}}{1 + e^{k_1 (x + v_1 t)} + e^{k_2 (x + v_2 t)}} \quad. \tag{19}$$

One then finds that

$$u_s = \frac{G_2}{u} = \frac{k_1 + k_2\, e^{(k_2 - k_1)(x + v_{1,2}\, t + \xi)}}{1 + e^{(k_2 - k_1)(x + v_{1,2}\, t + \xi)}} \qquad \left(\xi = \log\left[\frac{k_2}{k_1}\right]\right) \ . \tag{20}$$

$u(t,x)$ of Eq. (19) has three semi-infinite fronts, and vanishes asymptotically in a triangular wedge in the $x$-$t$ plane. $u_s$ of Eq. (20) has a single infinite ($M = 2$ semi-infinite fronts) with non-zero boundary values. It evolves along a characteristic line that is parallel to that of the inelastically generated front contained in $u(t,x)$ (see Eq. (17)). Both waves are shown in Fig. 1.

The case of the two-wave solution is indicative of what happens for general, $M > 1$: $u_s$ consists of $M$ semi-infinite fronts, which overlap with the inelastically generated fronts in $u$, the $M$-wave solution of the NF. This overlap extends along semi-infinite lines. This is why $R_{21}$, the obstacle to integrability, has the capacity to generate unbounded terms in the solution along each of the inelastically generated fronts, unless it is accounted for in a specific manner [14, 15].

The analysis in the case of the perturbed KdV equation follows similar steps. As an obstacle emerges only in second order [8-10], a perturbation through second order is included:

$$\begin{aligned}
w_t &= 6\, w\, w_x + w_{xxx} + \varepsilon\left(30\,\alpha_1\, w^2\, w_x + 10\,\alpha_2\, w\, w_{xxx} + 20\,\alpha_3\, w_x\, w_{xx} + \alpha_4\, w_{5x}\right) \\
&+ \varepsilon^2\begin{pmatrix} 140\,\beta_1\, w^3\, w_x + 70\,\beta_2\, w^2\, w_{xxx} + 280\,\beta_3\, w\, w_x\, w_{xx} \\ + 14\,\beta_4\, w\, w_{5x} + 70\,\beta_5\, w_x^{\ 3} + 42\,\beta_6\, w_x\, w_{4x} + 70\,\beta_7 w_{xx}\, w_{xxx} + \beta_8\, w_{7x} \end{pmatrix} + O\left(\varepsilon^3\right) \ .
\end{aligned} \tag{21}$$

The symmetries required through second order are [8-10, 17-19, 25]:

$$\begin{aligned}
S_1[u] &= u_x, \qquad G_1[u] = u \\
S_2[u] &= 6\, u\, u_x + u_{xxx}, \qquad G_2[u] = 3 u^2 + u_{xx} \\
S_3[u] &= 30\, u^2\, u_x + 10\, u\, u_{xxx} + 20\, u_x\, u_{xx} + u_{5x}, \qquad G_3[u] = 10\, u^3 + 10\, u\, u_{xx} + 5\, u_x^{\ 2} + u_{4x} \\
S_4[u] &= 140\, u^3\, u_x + 70\, u u_{xxx} + 280\, u\, u_x\, u_{xx} + 14\, u\, u_{5x} + 70\, u_x^{\ 3} + 42\, u_x\, u_{4x} + 70\, u_{xx}\, u_{xxx} + u_{7X}
\end{aligned} \tag{22}$$

Similarly to Eq. (3), the $O(\varepsilon)$ perturbation can be rewritten as:

$$\begin{aligned}
30\,\alpha_1\, w^2\, w_x + 10\,\alpha_2\, w\, w_{xxx} + 20\,\alpha_3\, w_x\, w_{xx} + \alpha_4\, w_{5x} &= \\
\tfrac{1}{3}\left(4\,\alpha_1 - 3\,\alpha_2 + 2\,\alpha_3\right) S_3[w] & \\
+ \tfrac{1}{3}\left(\alpha_1 - 2\,\alpha_2 - 2\,\alpha_3 + 3\,\alpha_4\right)\partial_x^{\ 4} S_1[w] + \tfrac{5}{3}\left(-\alpha_1 + \alpha_2\right)\partial_x^{\ 2} S_2[w] & \\
+ \tfrac{10}{3}\left(-\alpha_1 + 3\,\alpha_2 - 2\,\alpha_3\right) R_{21}[w] &
\end{aligned} \tag{23}$$

Thus, again, the quantity $R_{21}$, given by Eq. (4), appears in the perturbation to begin with. However, in the first-order calculation it does not constitute an obstacle to integrability. It can be eliminated by an appropriate term in the first-order NIT; the contribution of all $O(\varepsilon)$-terms in the perturbation can be accounted for in closed form, leading to a bounded first order correction, $u^{(1)}$ [8-10]. We note that unlike the case of the Burgers equation, here $R_{21}$ can be written as a gradient term.

The second-order perturbation cannot be written as a gradient term. A decomposition, similar to Eq. (8), can be found for the second-order perturbation, as well as for the driving term in the second-order homological equation. The freedom in the expansion allows one to reduce the obstacle to a term proportional to $u(t,x) \cdot R_{21}[u]$ [13, 14], which can be written as

$$u\,R_{21} = u\Big(S_2[u]\,G_1[u] - S_1[u]\,G_2[u]\Big) = u^3\,\partial_x\!\left(\frac{G_2[u]}{G_1[u]}\right) \quad . \tag{24}$$

Direct substitution shows that $G_2/G_1$ is constant ($= 4k^2$) for a single soliton solution of the KdV equation, given by

$$u(t,x) = \frac{2\,k^2}{\Big(\cosh\big\{k\big(x + 4\,k^2\,t\big)\big\}\Big)^2} \quad . \tag{25}$$

Hence, in the single-soliton case, Eq. (24) yields $R_{21} = 0$.

A multiple-soliton solution evolves asymptotically into a sum of well-separated single solitons [4, 5]. Hence, $G_2/G_1$ assumes constant values ($4 \cdot k_i^2$) in the vicinity of each single soliton, and changes from one constant value to the next somewhere between adjacent solitons. Thus, $G_2/G_1$ is composed of wave fronts. In Fig. 2, We show plots of $u(t,x)$ and $G_2/G_1$ for the case of a two-soliton solution, given by the Hirota formula [21]:

$$u(t,x) = 2\,\partial_x^2 \ln\!\left[ 1 + g_1(t,x) + g_2(t,x) + \frac{\big(k_1 - k_2\big)^2}{\big(k_1 + k_2\big)^2}\,g_1(t,x)\,g_2(t,x) \right] \quad .$$
$$\Big(g_i(t,x) = \exp\big\{2\,k_i\big(x + 4\,k_i^2\,t\big)\big\}\Big) \tag{26}$$

It is more instructive to consider $u_s$, defined as

$$u_s(t,x) \equiv \partial_x\!\left(\frac{G_2[u]}{G_1[u]}\right) \quad . \tag{27}$$

$u_s$ is also shown in Fig. 2 for the two-soliton case. Using Eq. (26), one finds that, away from the origin, $u_s$ asymptotes into a sum of two, well-separated, solitons that are generated through inelastic interactions of the Jost functions:

$$u_e \xrightarrow[|t| \to \infty]{} \frac{a_1}{\left(\cosh\left\{K_1\left(x+V_1 t\right)\right\}+\xi_1\right)^2} + \frac{a_2}{\left(\cosh\left\{K_2\left(x+V_2 t\right)\right\}+\xi_2\right)^2} \quad . \tag{28}$$

In Eq. (38), for $k_1 > k_2$, one has

$$a_1 = \text{sgn}(t) \cdot 2\left(k_1-k_2\right)\left(k_1+k_2\right)^2 \qquad a_2 = -\text{sgn}(t) \cdot 2\left(k_1+k_2\right)\left(k_1-k_2\right)^2 \quad , \tag{29}$$

$$\begin{aligned} K_1 &= k_1+k_2 \quad , \quad K_2 = k_1-k_2 \\ V_1 &= 4\left(k_1{}^2+k_2{}^2-k_1 k_2\right) \quad , \quad V_2 = 4\left(k_1{}^2+k_2{}^2+k_1 k_2\right) \end{aligned} \quad , \tag{30}$$

$$\xi_1 = \frac{\ln\left(\dfrac{k_1-k_2}{k_1+k_2}\right)+\sigma\ln\left(\dfrac{k_2}{k_1}\right)}{k_1+k_2} \quad , \quad \xi_2 = \frac{-\sigma\ln\left(\dfrac{k_2}{k_1}\right)}{k_1-k_2} \quad . \tag{31}$$

The second-order obstacle to integrability represents coupling between the KdV two-soliton state, and the spontaneously generated state, $u_s$, which also asymptotes into two well-separated solitons, but is *not* a solution of the KdV equation. The wave numbers and velocities of the new solitons are generated by inelastic interactions between the Jost functions from which the KdV solution is constructed. In addition, the amplitudes of the $u_s$-solitons have opposite signs, which they exchange upon a kinematically elastic collision. Finally, as the velocities of the new solitons are different from the velocities of the genuine KdV solitons, the characteristic lines in the *x-t* plane of the solitons in the two states are different. Therefore, the overlap between the multiplicative factors in Eq. (24) is confined to a small region around the origin, and the second-order canonical obstacle vanishes exponentially fast away from the origin. As a result, the obstacle does not have the capacity to generate unbounded terms in the expansion of the solution. However, its effect on the solution cannot be written in closed form as a differential polynomial in the zero-order solution $u$ [13, 14]. Extension to multiple-soliton solutions of the KdV equation is obvious.

<u>Figure Captions</u>

<u>Fig.1</u> Burgers equation: two-wave case. a) $u(t,x)$ (Eq. (18)); b) spontaneously generated wave (Eq. (19)). $k_1 = 5$, $k_2 = 3$.

<u>Fig. 2</u> KdV equation: two-wave case. a) $u(t,x)$ (Eq. (24)); b) spontaneously generated front, $G_2/u$; c) spontaneously generated sign-exchange solitons. $k_1 = 0.75$, $k_2 = 0.5$.

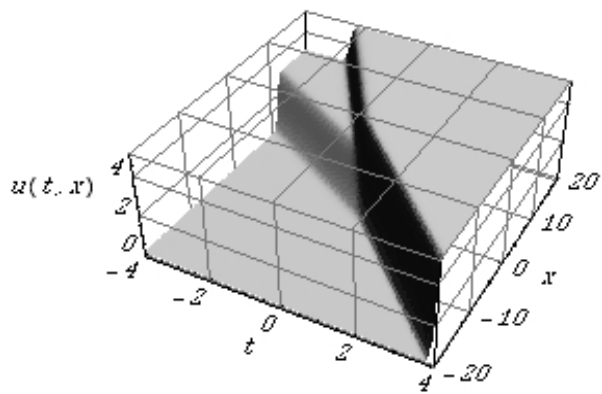

Fig.1a

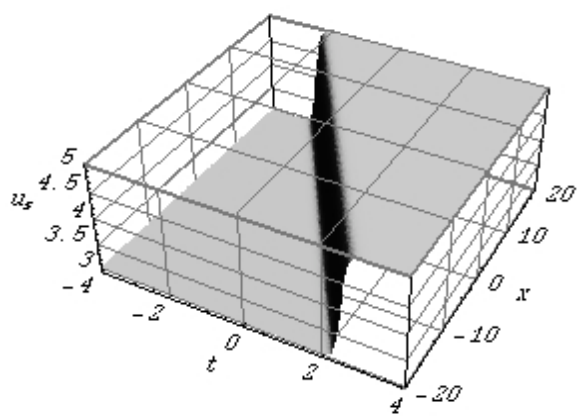

Fig. 1b

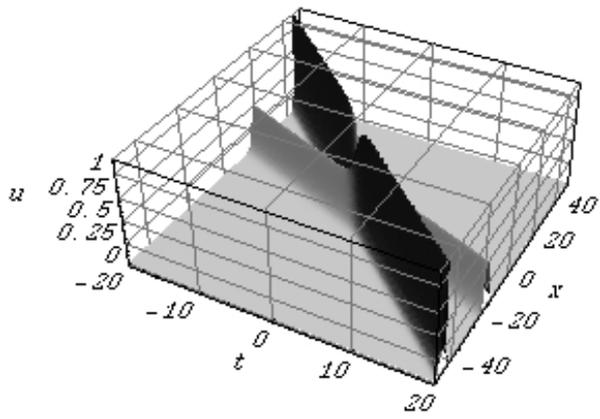

Fig.2a

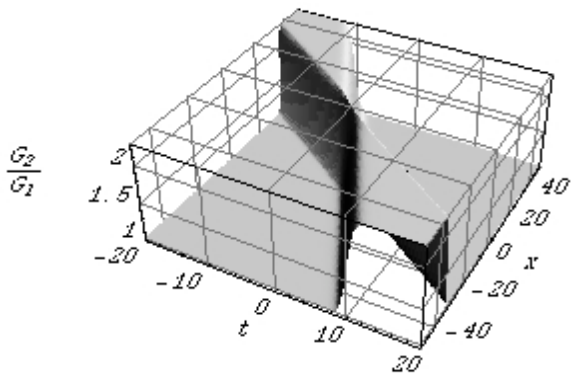

Fig.2b

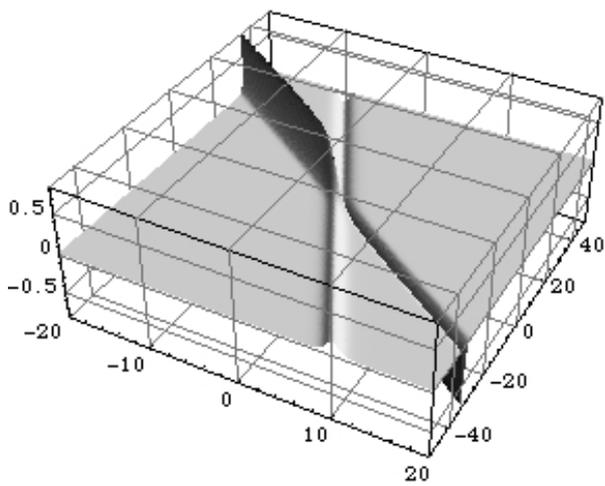

Fig.2c

-14-